\documentclass[a4paper,fleqn,usenatbib]{mnras}

\usepackage{newtxtext,newtxmath}

\usepackage[T1]{fontenc}
\usepackage{ae,aecompl}

\usepackage{graphicx}	
\usepackage{amsmath}	
\usepackage{amssymb}	
\usepackage{pdflscape}

\newcommand{\msun}{M$_{\odot}$} 
\newcommand{\kms}{\,km\,s$^{-1}$} 
\newcommand{\ha}{H$\alpha$} 
\newcommand{\hab}{H$\alpha_{\rm b}$} 
\newcommand{\han}{H$\alpha_{\rm n}$} 

\title[BHs revealed by photometric mass functions]{Hibernating black holes revealed by photometric mass functions}

\author[J. Casares]{Jorge Casares$^{1,2,3}$\thanks{Leverhulme Visiting Professor}\thanks{E-mail: jorge.casares@iac.es}\\
$^{1}$Instituto de Astrof\'isica de Canarias, 38205 La Laguna, Tenerife, Spain\\
$^{2}$Departamento de Astrof\'isica, Universidad de La Laguna, E-38206 La Laguna, Tenerife, Spain\\
$^{3}$Department of Physics, Astrophysics, University of Oxford,
Keble Road, Oxford OX1 3RH, UK \\
}

\date{Accepted XXX. Received YYY; in original form ZZZ}

\pubyear{2016}

\begin{document}
\label{firstpage}
\pagerange{\pageref{firstpage}--\pageref{lastpage}}
\maketitle

\begin{abstract}
We present a novel strategy to uncover the Galactic population of quiescent black holes (BHs). This is based on a new 
concept 
, the {\it photometric mass function} (PMF), which opens up the possibility of an efficient 
identification of dynamical BHs in large fields-of-view. This exploits the width of the disc \ha~emission line, combined 
with orbital period information.  
We here show that \ha~widths can be recovered using a combination of customized \ha~filters. By setting a width 
cut-off at 2200 \kms~we are able to cleanly remove other Galactic populations of \ha~emitters, including 
$\sim$99.9\% of cataclysmic variables (CVs). Only short period ($P_{\rm orb}<$2.1 h) eclipsing CVs and AGNs 
will contaminate the sample but these can be easily flagged through photometric variability 
and, in the latter case, also mid-IR colours. 
We also describe the strategy of a deep (r=22) Galactic plane survey based on the concept of  PMFs:  
{\it HAWKs}, the {\it HAlpha-Width Kilo-deg} Survey. We estimate that $\sim$800 deg$^2$ are required to unveil 
$\sim$50 new dynamical BHs, a three-fold 
improvement over the known population. For comparison, a century would be needed to produce 
an enlarged sample of 50 dynamical BHs from X-ray transients at the current discovery rate. 
\end{abstract}




\section{Introduction}
\label{intro}

The historic discovery of gravitational waves (GW) from merging BHs has undoubtedly opened a new era in 
astronomy \citep{abbott16a, abbott16b}. 
It means that BHs can now be detected without an electromagnetic counterpart. 
Also, the prediction of $\sim10^3$ new merging BHs to be discovered by Advanced LIGO in the next decade  
will allow for the first BH demographic studies across cosmological times \citep{abbott16c, sesana16, kovetz17, cholis17, elbert17}. 
Different formation scenarios, however, have been proposed to explain the GW events GW150914 and 
GW151226, including massive binaries at low metallicity, dynamical encounters in globular clusters or 
primordial BHs \citep{belczynski16, demink16, rodriguez16, coleman16, clesse17}. 
At present it is unclear which model can better accommodate the new GW BHs and it seems likely that a 
degenerate combination of multiple formation channels may be 
responsible for the entire population. In this context, the study 
of BHs in Galactic X-ray binaries remains extremely valuable because they present us with a homogeneous 
reference sample, drawn from a very specific formation channel in solar metallicity environment 
(e.g. \citealt{wang16} and references therein). This is precisely the scope of the current paper.
 
Galactic BHs are mostly discovered in X-ray transients (XRTs), a subclass of X-ray binaries which exhibit episodic 
outbursts triggered by accretion disc instabilities (see e.g. \citealt{belloni11}). These X-ray brightenings can be dramatic 
as they happen within a day or so, and can often outshine the brightest sources in the sky, making them easy to spot by 
X-ray monitoring satellites. But between these extraordinary outbursts, BH XRTs {\it hibernate} in a "quiescent" state, 
which can last for several decades or more, with typical X-ray luminosities below $\sim10^{32}$ erg s$^{-1}$. It is during these times 
 that it becomes possible to detect the faint low-mass donor star at optical/NIR wavelengths and exploit its kinematics to 
 constrain the masses of 
 the binary components. About 60 XRTs with suspected BHs (as indicated by their X-ray timing and spectral properties) 
 have been discovered since the dawn of X-ray astronomy but most of them faded below the detection threshold of 
 present instrumentation during their decay \citep{corral16}. 
Time-resolved spectroscopy in quiescence is required for dynamical confirmation (i.e. a mass function greater than 
$\sim$3 \msun) and this has only been achieved in $\sim$50 years for 17 BH transients \citep{casares-jonker14}.
For detailed population studies it is therefore critical to dramatically 
increase the number of dynamical BHs in X-ray binaries.  This requires a new research methodology rather than 
simply waiting for new transients to trigger an outburst, as this would take many decades to barely double the current 
number. 

In this regard, lessons can be learned from the study of supermassive BHs in quasars (QSOs) and Active 
Galactic Nuclei (AGNs). Nearby supermassive BHs are also weighed from the kinematics of central 
stars or, in exceptional circumstances, water radio masers.
However, in distant galaxies, when the BH radius of influence cannot be resolved, empirical 
scalings with galaxy properties (such as the $M_{\rm BH}$-$\sigma$ relation in the bulge and others based on 
reverberation mapping of the broad-line region, BLR) have to be applied (see \citealt{kormendy13} for a review of 
the different methods). In the most frequent form of reverberation-based techniques, the BH mass scales with the 
width of emission-lines formed in the BLR and the continuum luminosity, a proxy for the BLR size \citep{peterson04,vestergaard06}. 
Similarly, when the donor star is not detected in quiescent XRTs, scaling relations of spectral features with 
fundamental binary parameters may prove useful. 
For example, in \citet{casares16} (Paper II hereafter), we show that the ratio between the 
double-peak separation and the FWHM of the \ha~line, formed in the accretion disc, correlates with the binary mass ratio 
$q=M_2/M_1$ for $q\lesssim0.25$ (with $M_2$ and $M_1$ being the masses of the companion and BH, respectively). 
This stems from the fact that the double peak separation traces the outer disc truncation radius which, for these extreme 
mass ratios, is  determined by the 3:1 resonance tide with the companion star. The relation presented in Paper II can, 
for instance, provide a first estimate of the BH mass when $M_2$ is inferred from broad-band colors or the orbital 
period  (assuming the companion is a Roche-lobe filling Main Sequence star dominating the quiescent optical/NIR 
spectrum). 

Furthermore, in \citet{casares15} (hereafter Paper I) we find that the FWHM of the \ha~line scales with the projected 
velocity semi-amplitude of the companion star $K_2$, i.e.~\footnote{Note that a precursor of this equation was 
presented in \citet{warner73} (see also \citealt{jurcevic94}). These earlier versions were developed for the case 
of cataclysmic variables (i.e. interacting binaries with accreting white dwarfs) although, at variance with equation~(\ref{eq:fwhm-k2}), provide a scaling with the projected 
velocity of the primary star, $K_1$, as inferred from the motion of the wings of the H$\beta$ line.} 

\begin{equation}
K_2=0.233~FWHM 
\label{eq:fwhm-k2}
\end{equation}

\noindent
This equation suggests a fundamental scaling between the dynamics of the gas in the accretion disc and the companion 
star. It should be mentioned that the slope of the FWHM-K2 correlation does contain  
a weak dependence on $q$ (see eqs. 5-6  in Paper I); for example, cataclysmic 
variables above the 
period gap display a  $\sim$27\% flatter slope than shown by equation~(\ref{eq:fwhm-k2}). In any case, the FWHM-K2 correlation deploys very tightly for BH XRTs 
because their mass ratios cluster within a very narrow range centered at $q\sim0.05$. 

As a matter of fact, equation~(\ref{eq:fwhm-k2}) proves very useful to study faint XRTs: it simply requires 
resolving the width of the 
\ha~line (usually the strongest feature in the spectrum) rather than measuring the doppler shifts of weak 
absorption lines from the donor star in phased-resolved spectra. Therefore, if the orbital period is known (e.g. through 
photometric variability) 
then mass functions can be extracted from single-epoch spectroscopy. This effectively allows extending dynamical 
studies to XRTs $\sim2$ magnitudes fainter than is currently possible, anticipating what 
40-m class telescopes can achieve in the next decade using traditional techniques. 
Examples of this strategy are presented in 
\citet{zurita15}, \citet{mata15} and Torres et al. (in preparation), see also Paper II. 

In addition, Paper I introduces a new concept, the {\it photometric mass function} (PMF), expressed by  

\begin{equation}
\left(\frac{PMF}{M_{\odot}}\right)\equiv1.3\times10^{-9}~\left(\frac{P_{\rm orb}}{d}\right)~
\left(\frac{FWHM}{km~s^{-1}}\right)^3 = \left(\frac{M_1}{M_{\odot}}\right)\frac{\sin^3 i}{\left(1+q\right)^2} 
\label{eq:pmf}
\end{equation}

\noindent
Here PMF refers to the case when both \ha~widths ( i.e. FWHM) and orbital periods ($P_{\rm orb}$) are 
derived {\it photometrically}. 
Since mass functions provide a lower limit to the mass of the accreting star, PMFs (as delivered by imaging techniques) 
can be exploited to discover quiescent BHs very efficiently in large fields of view. 

In this paper we present a proof-of-concept on how \ha~FWHMs can be measured through photometry  
using custom interference filters, and how this can be exploited to filter out 
contaminating sources that plague classic \ha~surveys (Section~\ref{sec:filters}).     
In Section~\ref{sec:survey} we propose a survey strategy to identify hibernating BHs from \ha~widths and 
PMF selection. Finally, in Sections~\ref{sec:discussion} and~\ref{sec:conclusions} we discuss our results and summarize the conclusions. 
We envision that this new methodology has the potential for boosting the statistics of dynamical 
BHs by an order of magnitude within only a few years.  

\section{A Photometric System to Measure \ha~Widths}
\label{sec:filters}

Dormant BH XRTs are very difficult to identify because of their extremely low accretion luminosities.  In turn, the 
companion (typically a K-type star) dominates the optical spectrum, thereby effectively disguising quiescent XRTs 
amongst the myriad of field stars. Only the presence of superimposed broad emission lines from the accretion disc, most notably 
\ha, can betray their presence. In fact, current 
\ha~surveys of the Galactic plane, 
such as the {\it SuperCOSMOS H-alpha Survey} (SSH, \citealt{parker05}) or the {\it INT Photometric  
\ha~Survey of the Northern Galactic Plane} (IPHAS, \citealt{drew05}) 
might contain quiescent BHs but, unfortunately, they are vastly outnumbered by other populations of \ha~emitters 
such as symbiotic binaries, cataclysmic variables (CVs), Be stars or planetary nebulae.  
For example, among the $\sim10^4$ \ha~sources contained in IPHAS 
\citep{drew05, witham08} no BH candidates have yet been produced. 
Another approach, followed by the {\it Galactic Bulge Survey} (GBS) 
relies mainly on cross-matching \ha~emitting objects  
with weak X-ray sources from a shallow Chandra survey of the Galactic Bulge \citep{jonker11,jonker14}.  
About 25 new quiescent BHs were predicted by GBs but only intervening CVs (mostly magnetic), W UMa or 
coronally active stars have been securely identified so far (e.g. \citealt{torres14,wevers17}).

\begin{figure}
	\includegraphics[angle=0,width=\columnwidth]{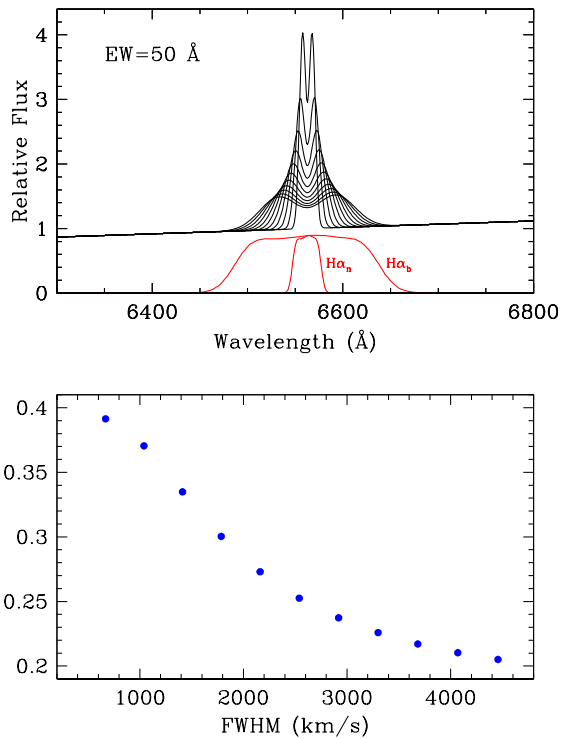}
    \caption{Top: Synthetic double-peaked \ha~spectra with FWHM=670-4450 \kms, together with the transmission 
    curves of our simulated \ha~filters. Bottom: Variation of the filter's flux ratio versus FWHM. The flux ratio is 
    computed through the convolution of the synthetic spectra with the \ha~filters.}
    \label{fig:fig1}
\end{figure}

Here we propose a new route 
that exploits the width of the \ha~line as a tracer of the strong gravitational field 
exclusive to BHs. From the compilation presented in Paper I we note that quiescent BHs typically possess 
FWHM$\gtrsim$1000 km s$^{-1}$, i.e.  much larger than observed in most other populations of \ha~emitters. 
Therefore, in order to flush out potential contaminants it would be of great interest to develop a photometric system 
tailored to measure \ha~line widths. 
In principle, this could be achieved with just a pair of \ha~interference filters: one sufficiently broad to contain the 
entire line flux (hereafter called \ha~{\it broad}, \hab) and another filter matched 
to sample only the line core (\ha~{\it narrow}, \han). 
To test this idea we have simulated a pair of \hab~ and \han~filters by scaling filter \#197 from the Wide Field 
Camera (WFC) on the 2.5m Isaac Newton Telescope (INT), which has an effective FWHM of 
95 \AA~\footnote{The transmission curve of this filter is available from 
http://catserver.ing.iac.es/filter/filtercurve.php?format=txt\&filter=197}. The two simulated 
filters have been shifted to the \ha~rest wavelength i.e. 6562.76 \AA. 
For the \hab~filter we take a bandwidth of 150 \AA~so as to encompass the entire flux of the broadest BH \ha~line 
currently known i.e. that of Swift J1357-0933 (Paper I). Regarding the \han~filter we start by adopting a 
bandwidth of 33 \AA~(=1508 km $s^{-1}$). 
In addition, we have produced a set of synthetic "interacting binary" spectra 
consisting of double-peaked \ha~profiles with fixed EW=50 \AA~(we take EWs of emission lines as positive 
henceforth) and FWHM=670-4450 km s$^{-1}$ sitting on a reddened linear continuum (Appendix~\ref{ap:profiles}). 
In the top panel of Figure~\ref{fig:fig1} we present the synthetic spectra, together with the transmission curves of 
the simulated filters. 

\begin{figure}
	\includegraphics[angle=0,width=\columnwidth]{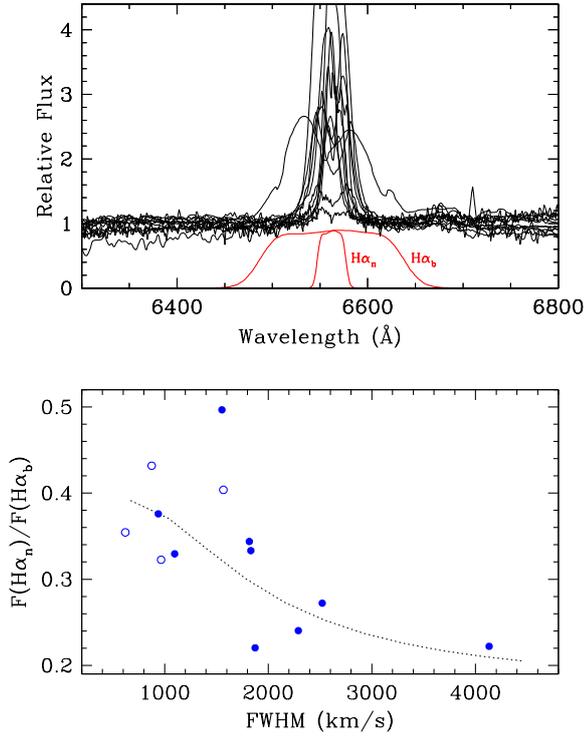}
    \caption{Same as Figure~\ref{fig:fig1} but for a sample of 9 BH (filled circles) and 4 CV (open circles) 
    real spectra. Error bars in the lower panel are smaller than the symbol size and are not displayed.  
    The dotted line indicates the trend  derived from synthetic spectra, with EW=50 \AA.} 
    \label{fig:fig2}
\end{figure}

The synthetic spectra were convolved with the transmission curves of the \hab~and \han~filters and  
 integrated over wavelength to derive associated fluxes (Appendix~\ref{ap:profiles}). 
 Here and henceforth, we sample the filter's response curves and the spectra to a common wavelength scale 
 of 1 \AA~pix$^{-1}$. 
The bottom panel in Figure~\ref{fig:fig1} displays the ratio of fluxes obtained from the pair of \ha~filters as a 
function of line FWHM. The latter was  derived through fitting Gaussian functions to the synthetic profiles (see 
Appendix~\ref{ap:profiles}). The smooth curve indicates that FWHM values can indeed be recovered 
from the combined fluxes provided by the simulated filters. We subsequently applied the same exercise to a sample 
of flux-calibrated spectra of BHs and CVs from the compilation presented in Paper I~\footnote{The sample contains 
9 BHs (V404 Cyg, BW Cir, GRO J0422+320,  XTE J1650-50, N. Oph 77, A0620-00, GS2000+25, XTE J1118+480, 
Swift J1357-0933) and 4 CVs (CH UMa, SS Aur, SS Cyg, U Gem) covering a wide range of FWHMs between 
600 and 4200 \kms. Spectral fluxes have been normalized to the continuum level at \ha.} and the result is presented 
in Figure~\ref{fig:fig2}. This time the smooth trend is replaced by significant scatter, caused by large 
variations in line EW from system to system. This can be interpreted as the effect of different underlying continua  
diluting the contribution of the line to the total flux. In other words, a broad emission line sitting on a strong 
continuum can mimic the same flux ratio as a narrow line over a weak continuum.   
Therefore, in order to measure accurate line widths it is crucial to know the EW of the line beforehand and 
this can be accomplished with an extra observation using a broader filter (r-band hereafter), also centered at \ha.  

\subsection{Photometric EWs}
\label{sec:pew}

\begin{figure}
	\includegraphics[angle=0,width=\columnwidth]{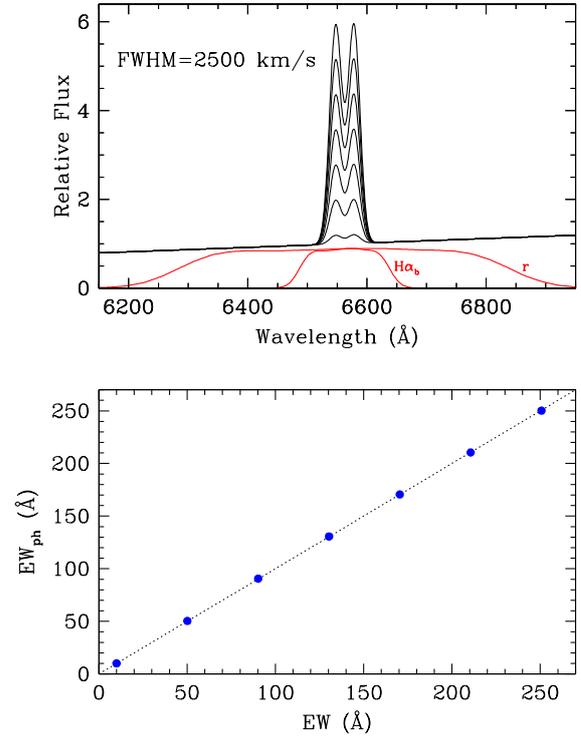}
    \caption{Top: Synthetic double-peaked \ha~spectra with FWHM=2500 \kms~ and different EWs in the range 10-250 
    \AA, together with the transmission curves of our simulated \hab~and r-band filters. 
    Bottom: Photometric EWs (EW$_{\rm ph}$), as extracted from equation ~\ref{eq:pew} with $C_{1}=1.155$, 
    versus model EWs. The dotted line represents EW$_{\rm ph}=$EW.} 
    \label{fig:fig3}
\end{figure}

It is easy to prove that the EW of an \ha~line can be approximated by the following equation   

\begin{equation}
EW_{ph}=C_{1}~\frac{W_{r} \times \left(\frac{F_{H\alpha_b}}{F_r}\right)- W_{H\alpha_b}}{1-\left(\frac{F_{H\alpha_b}}{F_r}\right)} 
\label{eq:pew}
\end{equation}

\noindent
where $W_{H\alpha_b}$ and $W_{r}$ stand for the equivalent widths 
of the \hab~and r-band filters, $F_{H\alpha_b}$ and 
$F_{\rm r}$ are the fluxes measured by the filters and $C_{1}$ a calibration constant 
(see Appendix~\ref{ap:pew}). We call this expression $EW_{ph}$ or {\it photometric} 
EW i.e. the EW as derived from observations with the two photometric filters.  
Here again we have simulated the r-band filter by scaling filter \#197 of the WFC to a bandwidth of 550 \AA~and 
an effective central wavelength of 6562.76 \AA. The width of the r-band filter was adopted to avoid the strong telluric 
\ion{O}{I} emission line at $\lambda$6300 and the B-band absorption at $\sim6860$ \AA, which may affect the 
accuracy of photometric measurements. In the specific case of our filters we find $W_{H\alpha_b}=132.7$ 
\AA~and $W_{r}=487.0$ \AA~through integration of the filter's transmission curves. 
In order to determine $C_1$ we created a set of synthetic X-ray binary spectra with FWHM=2500 \kms and 
EWs in the range 10-250 \AA. These were subsequently convolved with the transmission curves of the filters to 
yield $F_{H\alpha_b}$ and $F_{\rm r}$ fluxes. EW$_{ph}$ values are then computed through equation~\ref{eq:pew} 
and compared to model EWs. A linear fit to the sequence of EW$_{ph}$ and model EW values leads to 
$C_{1}=1.155$. The results are displayed in Figure~\ref{fig:fig3}. 

\subsection{Photometric FWHMs}
\label{sec:pw}

\begin{figure}
	\includegraphics[angle=0,width=\columnwidth]{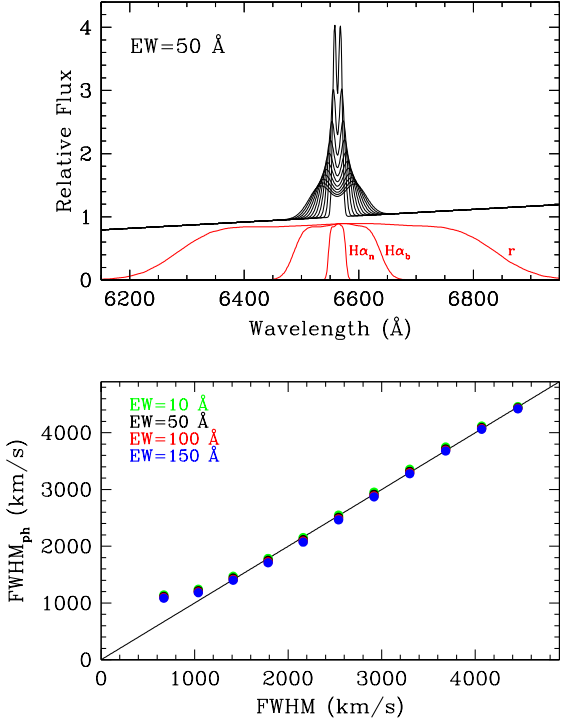}
    \caption{Top: Example synthetic double-peaked \ha~spectra with EW=50 \AA~and FWHM in the range 670-4450 
    \kms, together with the transmission curves of our simulated \han, \hab~and r-band filters. 
    Bottom: Photometric FWHMs (FWHM$_{\rm ph}$), as provided by equation ~\ref{eq:pw} with $C_{2}=0.826$, 
    versus FWHMs measured  through a Gaussian fit for different model EWs in the range 10-150 \AA. The dotted line 
     represents FWHM$_{\rm ph}=$FWHM. Note that only spectra with FWHM$\leq1200$ \kms~deviate from the line 
      because their widths are smaller than the bandwidth of the narrow \ha~filter.} 
    \label{fig:fig4}
\end{figure}

Once EW$_{\rm ph}$ is determined, it is then possible to extract the FWHM of the line through the expression 

\begin{equation}
FWHM_{ph}=C_{2}~\frac{EW_{ph}}{\left(\frac{EW_{ph}+W_{H\alpha_b}}{W_{H\alpha_n}}\right)\times\left(\frac{F_{H\alpha_n}}{F_{H\alpha_b}} \right)-1} 
\label{eq:pw}
\end{equation}

\noindent
where $W_{H\alpha_n}$ is the equivalent width of the \han~filter ($W_{H\alpha_n}=26.3$ \AA~in our case), 
$F_{H\alpha_n}$ the flux measured through that filter and $C_{2}$ a calibration constant (see Appendix~\ref{ap:pw}). 
In the same way as before, $C_2$ has been obtained by comparing FWHM values of synthetic spectra 
(with fixed EW=50 \AA, see top panel in Figure~\ref{fig:fig4}) with FWHM$_{\rm ph}$ values computed from 
equation~\ref{eq:pw}, resulting in $C_2=0.826$.  The bottom panel of Figure~\ref{fig:fig4} displays the behaviour of 
FWHM$_{\rm ph}$ versus model FWHMs for a grid of synthetic spectra with EW=10-150 \AA~and FWHM=670-4450 
\kms. The figure demonstrates that FWHMs are now uniquely determined through equation~\ref{eq:pw}, 
irrespectively of the line EW. Only at small widths $\le1200$ \kms~do the FWHM$_{ph}$ values 
deviate significantly from model FWHMs but this is only because we are limited by the bandwidth of the narrow 
\ha~filter which determines our effective resolution in width. 

\begin{figure}
	\includegraphics[angle=0,width=\columnwidth]{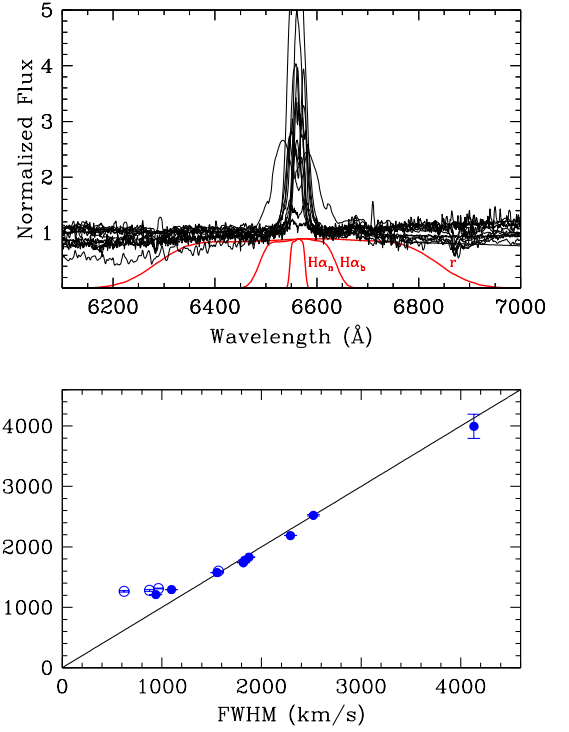}
    \caption{Same as Figure~\ref{fig:fig4} but for a sample of 9 BHs (filled circles) and 4 CVs (open circles). 
    Error bars in the lower panel are smaller than the symbol size in all but one case. Photometric widths 
    recover real FWHMs to better than 5\% for FWHM$\ge1200$ \kms. }
    \label{fig:fig5}
\end{figure}

A final test is performed with our previous sample of real BH and CV spectra and the results are presented in 
Figure~\ref{fig:fig5}. In contrast to Figure~\ref{fig:fig2}, the variations in line EW are now accounted for, 
allowing us to recover line widths at $<$5\% accuracy for $FWHM\ge1200$ \kms.  
Here 5\% indicates the largest fractional difference between the true data FWHM and the 
FWHM$_{\rm ph}$ values provided by eq. 4.
Therefore, photometric observations with a combination of only three filters (broadband r plus 
two narrowband \ha~filters) are sufficient to determine the width (and equivalent width) of 
the \ha~line in quiescent BH XRTs. 

\subsection{Width Cut-off for BH selection}
\label{sec:cutoff}

Having proved that line FWHMs can be reliably measured with our photometric system we now need to 
define an optimal width cut-off to efficiently select quiescent BHs. This is an important issue as it 
determines the fraction of other \ha~emitting objects that will be rejected. Larger cut-off widths 
will filter out narrow \ha~emitters but also BHs at lower inclinations and, therefore, a compromise 
must be devised. For example, a cut-off at FWHM$\sim$1500 \kms~will instantly clean out most 
potential contaminants such as T Tau, Be stars, chromospherically active stars or planetary nebulae. 
Only CVs at moderately high inclinations can produce \ha~lines this broad because of the deep potential wells 
of their accreting white dwarfs. However, with a Galactic density of $\sim10^4$ kpc$^{-3}$ (\citealt{pretorius12}, 
see also \citealt{burenin16} for an update based on recent constraints on the X-ray luminosity 
function) CVs are a factor $\approx10^3$ more numerous than BH XRTs (e.g. \citealt{corral16}) 
and most likely will dominate the Galactic population of \ha~contaminants, even at these large widths. So it is 
important to set a more stringent width cut-off in order to optimize the rejection of potentially contaminating CVs. 

\begin{figure}
	\includegraphics[angle=0,width=\columnwidth]{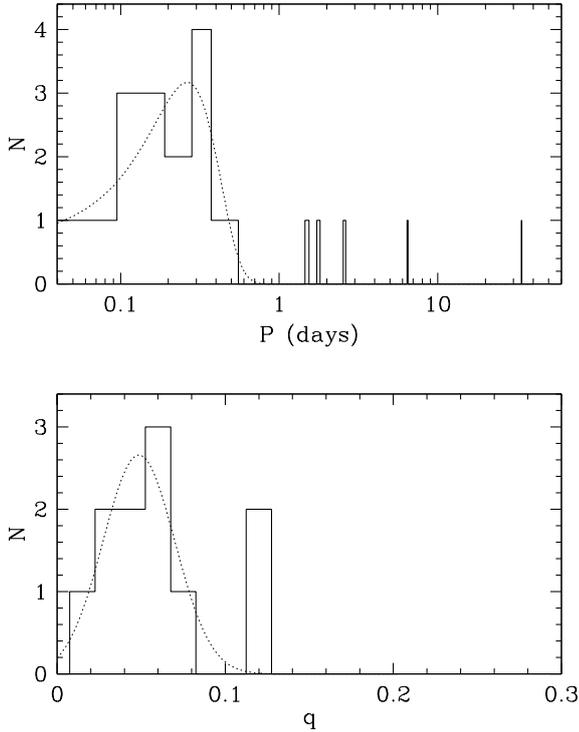}
    \caption{Observed distribution of orbital periods (top) and mass ratios (bottom) for BH XRTs. The dotted line 
    represents the best Gaussian fit. 
    } 
    \label{fig:fig6}
\end{figure}

To investigate this issue we have run Monte Carlo simulations (10$^5$ trials) of the expected distribution of FWHMs 
for the two binary populations. In the case of BH XRTs we have computed FWHMs from equation~\ref{eq:pmf}, 
assuming an isotropic distribution of inclinations i.e. $\langle \cos i \rangle=0.5$. 
Following \citet{ozel10} we have drawn 
BH masses from a normal distribution with mean 7.8 \msun~and $\sigma=1.2$ \msun. $P_{\rm orb}$ and  $q$ 
values are also obtained from normal distributions with $\langle P_{\rm orb} \rangle$=0.287 d, 
$\sigma(P_{\rm orb})=0.14$, $\langle q \rangle$=0.05 and $\sigma(q)=0.02$.  
These have been parametrized from Gaussian fits to the observed distributions  
of $P_{\rm orb}$  and mass ratios in BH XRTs (Figure~\ref{fig:fig6}), as listed in \citet{casares-jonker14} and Paper II. 
The sample also includes the orbital periods of Swift J1753.5-0127 \citep{zurita08}, Swift J1357.2-0933 \citep{corral13} 
and MAXI J1650-152 \citep{kuulkers13}, the shortest currently known. Note that BH XRTs with 
intermediate-mass donor stars (i.e.  XTE J1819.3-2525, GRO J1655-40 and  4U 1543-475, all with 
$P_{\rm orb}\simeq1-3$ d and $q\gtrsim0.3$) are not considered because of the lack of \ha~emission in their 
spectra. 
Finally, the area of the FWHM probability distribution function (PDF)  
has been scaled by a factor 1000 to account for the larger size of the Galactic population of CVs relative to 
BH XRTs.  

For the case of CVs, we start by assuming the $P_{\rm orb}$ distribution of the 1146 CVs listed in the 7th edition of 
the \citet{ritter03} catalogue (Release 7.21). This contains the sample of intrinsically faint short-period CVs 
discovered by Sloan \citep{gansicke09} and will be referred to as  
{\it model CV-1} henceforth.  
The ultracompact AM CVn binaries are excluded because they possess degenerate donor stars and, 
therefore, they lack Balmer emission lines.  Since mass ratios are known to increase with $P_{\rm orb}$ 
(e.g. \citealt{knigge11}), we adopted an exponential dependence of the form 
$q=0.731-11.55\times {\rm e}^{-\left(P_{orb}+0.388\right)/0.154}$, 
derived through a least-square fit to 117 pairs of $P_{\rm orb}-q$ values from \citet{ritter03}.
White dwarf masses are drawn from a normal distribution with mean=0.83 \msun~and $\sigma=0.23$ \msun~\citep{zorotovic11}, while binary inclinations are also assumed to be randomly distributed.
As before, $P_{\rm orb}$, $q$, $M_1$ and $cos i$ values are used as inputs to 
equation~\ref{eq:pmf}, with one proviso: because of the wide range of $q$ values for CVs we do not 
assume the FWHM-K$_2$ scaling from equation~\ref{eq:fwhm-k2} but the more general expressions given by 
equations 5 and 6 in Paper I, with $\alpha=0.42$. 

\begin{figure}
	\includegraphics[angle=0,width=\columnwidth]{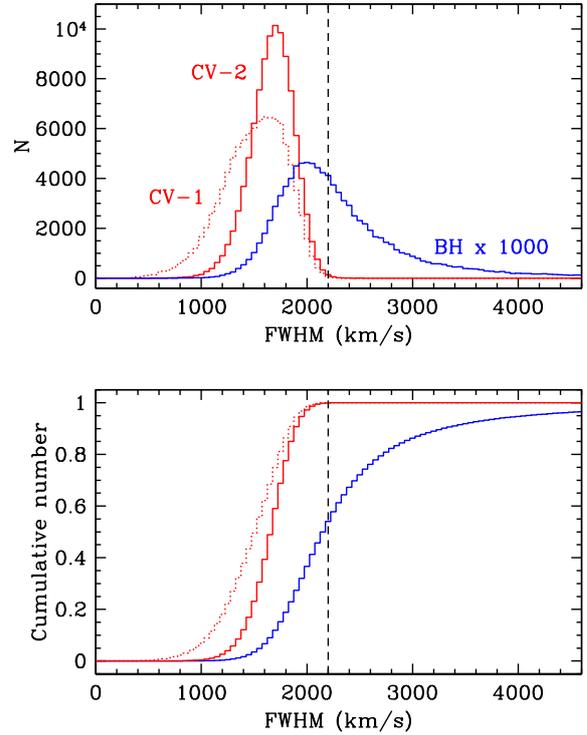}
    \caption{Top: Montecarlo PDFs of FWHMs for BHs and CVs (models 1 and 2). For the sake of display, the BH 
     distribution has been enlarged by a factor 1000. The dashed vertical line at FWHM=2200 \kms~marks our selected 
      cut-off width. Bottom: Normalized cumulative distributions of the number of CVs and BHs as a function of FWHM. 
      A width cut-off at FWHM=2200 \kms~allows selecting $\sim$46\% of BHs while rejecting $\sim$99.9\% of CVs.} 
    \label{fig:fig7}
\end{figure}

The number of CVs below the period gap in model CV-1 
is a factor $\sim$2 larger than those above the gap. This 
fraction, however, is most likely biased low by selection effects \citep{gansicke05} and hence  
we decided to compute a further PDF distribution where, following 
the prediction of standard CV population models,  we have incremented the number of CVs below the gap to 
98\% of the total population \citep{kolb93, howell01, knigge11} . 
We call this {\it model CV-2} 
and take it as a more realistic representation of the true Galactic population of CVs.   
Figure~\ref{fig:fig7} presents the PDFs of the FWHM and their cumulative distribution functions 
for BHs and the two CV models, while Table~\ref{tab:tab1} lists the percentage of the total populations selected 
by specific FWHM cut-off values. Because of the larger number of short period CVs in model CV-2, the fraction 
rejected by a given cut-off will be lower than in model CV-1. Since CVs are $\approx10^3$  times more 
abundant than BHs, we tentatively adopt FWHM$>$2200 \kms~as our optimal cut-off value. According to the  
simulation, this cut-off would allow the rejection of $\sim$99.9\%~of CVs while retaining $\sim$46\%~of BHs i.e. 
under the assumption of equal absolute magnitudes and Galactic distributions it would select $\approx$2 CVs 
per BH discovered.   

\subsection{The \ha~Colour Diagram}
\label{sec:colour}

\begin{figure}
	\includegraphics[angle=-90,width=\columnwidth]{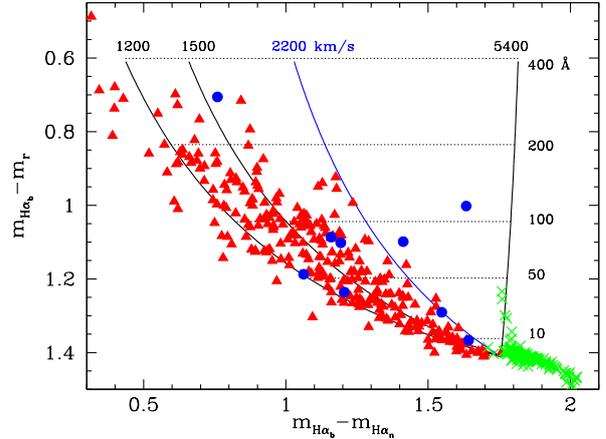}
    \caption{The \ha~colour-colour diagram. Dotted horizontal lines represent constant EWs in the range 10-400 
    \AA~while vertically curved continuous lines indicate constant FWHM in the range 1200-5400 \kms. 
    The blue line at  FWHM=2200 \kms~marks our favored width cut-off for efficient BH selection. 
    SDSS CVs are located by red triangles, BH XRTs by blue dots and spectral type standards of luminosity class 
    I, III and V by green stars.      } 
    \label{fig:fig8}
\end{figure}

For practical purposes, 
Figure ~\ref{fig:fig8} presents a colour-colour diagram constructed using the simulated filters described in  
Sections~\ref{sec:pew} and~\ref{sec:pw}. We call this the 
{\it \ha~Colour Diagram} and contains the 
 information on FWHM and EW of \ha~emitting stars. 
To guide the eye, lines of constant FWHM 
and EW are overplotted, as calculated through convolution of synthetic double-peaked profiles with our filter 
transmission curves (see Appendix~\ref{ap:profiles}). 
A line of maximum FWHM has been drawn at 5400 \kms, based on the extreme case of 
a 12 \msun~BH with a 0.07 \msun~companion star (right at the H-burning limit) in a 1.37 hr orbit, 
seen edge-on ($i=90^{\circ}$). This orbital period has been derived assuming that the companion star fills its 
Roche lobe and obeys the semi-empirical Mass-Radius relations from \citet{knigge11}.    
Our width cut-off at FWHM=2200 \kms~is indicated by the blue continuous line. 
The optimum region for BH detection in the \ha~Colour Diagram is hence restricted between the FWHM=2200 
and 5400 \kms~lines.  
For comparison, we have plotted in blue circles synthetic colours of our sample of dynamical BHs while 
285 bona-fide CVs, selected from Sloan DR7 \citep{szkody11}, have been marked by red triangles. 
Because of the survey depth and broad colour selection cuts Sloan CVs provide the least 
biased sample obtained so far (see \citealt{gansicke05}) and hence we take these as the 
best representation of the field CV population that one might expect to find. 
Finally, synthetic colours of a grid of O-M Main Sequence, Giants and Supergiant stars from 
\citet{jacoby84} are also plotted as green crosses. 

We note that \ha~emitters are cleanly segregated in this diagram from 
non-\ha~emitting field stars. The latter tend to cluster at EW$\approx$0, i.e. near the focus 
of lines of constant FWHM, with some spreading along two characteristic directions: 
(1) a tail towards EW$<0$ and large FWHMs, produced by A-B stars with broad \ha~absorptions (also a locus for 
isolated DA 
white dwarfs and CVs in outburst) and (2) a vertical stream at FWHM$\sim$5400 \kms~and 
EW$\sim0-50$ \AA, caused by M-type stars (mostly Giants and Supergiants) with deep molecular bands. 
It should be noted that, because our three filters are relatively narrow and have a common central wavelength, 
this diagram is insensitive  
to reddening. Therefore, stellar positions do not depend on extinction and both EWs and FWHMs of \ha~lines 
are uniquely determined, irrespectively of the slope of the underlying continuum. 
This represents a clear advantage over other \ha~surveys based on broad band 
colours, where EWs often become degenerate with interstellar extinction and the spectral energy distribution of the 
star (see e.g. \citealt{drew05}). 

\begin{figure}
	\includegraphics[angle=-90,width=\columnwidth]{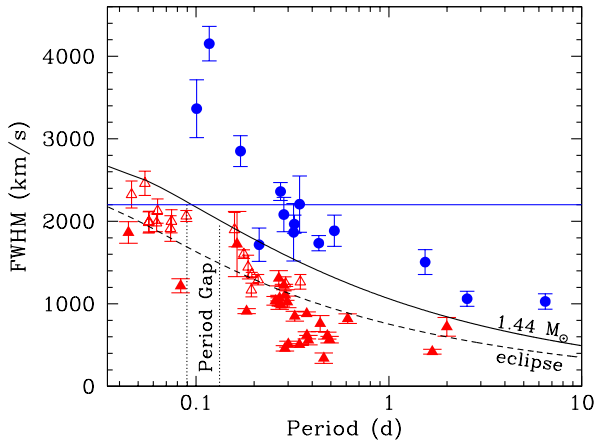}
    \caption{FWHM versus P$_{\rm orb}$ for a sample of BH XRTs (filled blue circles) and CVs (red triangles). 
    Eclipsing CVs are indicated by open red triangles. The black solid line indicates a hard upper limit in FWHM for 
    CVs, set by the Chandrasekhar mass, while the black dashed line represents a lower limit for eclipsing CVs. 
    The blue horizontal line marks our FWHM cut-off and the vertical dotted lines the edges of the period gap for CVs. 
    Only eclipsing CVs with orbital periods under the gap can produce \ha~lines with FWHM$>$2200 \kms.  
     } 
    \label{fig:fig9}
\end{figure}

The \ha~colour diagram proves that the great majority of Sloan CVs are nicely rejected by the FWHM=2200 
\kms~cut-off, with only three exceptions: DV UMa, IY UMa and SDSS J122740.83+513925.0. These three 
CVs are eclipsing and have orbital periods in the range 1.52-2.06 hr. 
As a matter of fact, only CVs with very short orbital periods and 
extreme inclinations (thus eclipsing) are able to produce \ha~lines broader than FWHM$\ge$2200 \kms. This is   
exemplified by Figure~\ref{fig:fig9}, where we display FWHMs of \ha~lines versus $P_{\rm orb}$ for the sample of 
dynamical BHs (blue circles) and CVs (red solid triangles) presented in Paper I. Eclipsing CVs are distinguished 
by red open triangles. A recent determination of the FWHM of \ha~for MAXI J1659-152 has also been 
included (Torres et al. in preparation).  
The continuous black line represents a hard upper limit on FWHM for CVs, as obtained from the 
PMF expression (eq.~\ref{eq:pmf}), 
assuming maximum inclination $i=90^{\circ}$ and the Chandrasekhar mass limit, combined with the 
$P_{\rm orb}-q$ dependence derived in Section~\ref{sec:cutoff} and equations 5 and 6 from Paper I. 
In addition, we draw in dashed line style a lower limit for eclipses of the accretion disc. This has been obtained in the same 
way as before albeit for a typical white dwarf mass of 0.83 \msun~\citep{zorotovic11} and the eclipse condition 

\begin{equation}
\cos i\ge \left[\left(\frac{R_{d}}{R_{L1}}\right)+\frac{1-\left(\frac{R_{d}}{R_{L1}}\right)}{0.462}\times~\left(\frac{1+q}{q}\right)^{1/3}\right]^{-1} 
\label{eq:eclipse}
\end{equation}

\noindent
with $R_{d}/R_{L1}=0.5$ i.e. the accretion disc radius extending to 0.5 times the primary Roche lobe 
(e.g. \citealt{marsh94}). This expression is derived from simple geometry using Paczy\'nski's approximation 
for the radius of the companion's Roche lobe \citep{paczynski71}. 
The vertical dotted lines mark the extremes of the period gap, between 2.15 hr and 3.18 hr. 
Figure~\ref{fig:fig9} demonstrates that only eclipsing CVs with $P_{\rm orb}$ under the gap are able to  
produce \ha~lines broader than 2200 \kms. Therefore, those CVs not rejected by our width cut-off will be easily 
identified from the presence of $\sim$2-3 mag optical eclipses in light curves spanning $\lesssim$2 hr. 
In addition, most will be members of the WZ Sge-class (see \citealt{kato15}) and, thus, can also be 
flagged by the blue excess associated to the dominant white dwarf spectrum.  
We note in passing that such eclipsing CVs will be of great scientific value in their own right, in particular for 
computing accurate white dwarf masses and evolutionary studies (e.g. \citealt{littlefair08}). 

\subsection{Other possible outliers}
\label{sec:contaminants}

 While high-inclination short-period CVs represent our main concern, some contamination may also be expected from 
other stellar populations. For example, broad molecular bands in late M-type stars (see Figure~\ref{fig:fig8}) 
and S-type symbiotic binaries \citep{munari02} can mimic very broad \ha~emission lines with FWHM$\approx$5400 \kms. 
These could be confused with extreme BHs but are readily spotted by their extreme red colours in optical/NIR 
broadband photometry. D-type symbiotics, on the other hand, pose no hazard as their nebular-type spectra 
typically have very large EWs $\gtrsim$1000 \AA.  

A small fraction of WN-type Wolf-Rayet stars \citep{smith96} 
can certainly produce very broad emission lines (\ha, \ion{He}{I}, \ion{He}{ii}) 
within the spectral band covered by our narrow-band filters  
but they are very scarce and tightly confined to star forming clusters,  
not to be targeted by our proposed survey (see Section~\ref{sec:requirements}). 
Likewise, QSOs and AGNs (Seyfert 1s) at specific redshifts can bring 
broad emission lines into our photometric band. Contamination from \ha~and \ion{N}{iii} $\lambda$6548 
will be produced at $z<0.01$ although these nearby AGNs are a trifle and will be immediately recognized 
as extended (non-stellar) objects. 
Further contamination from H$\beta$, H$\gamma$, \ion{Mg}{ii}$\lambda$2798, 
\ion{C}{iii}$\lambda$1909, \ion{C}{iv}$\lambda$1459 and Ly$\alpha$ is expected at redshifts 
z=0.34$-$0.36, 0.50$-$0.52, 1.33$-$1.36, 2.41$-$2.47, 3.44$-$3.56 and 4.3$-$4.5, respectively. 
Scaling from the 13th edition of the Veron Catalogue of QSO \& AGNs \citep{veron10} we estimate   
an incidence of $\sim$0.5 deg$^{-2}$  (66\% from \ion{Mg}{ii} and \ion{C}{iii}) for magnitudes $V\leq$22. 
However, given the observed distribution of line widths \citep{puchnarewicz97} only about  
$\sim$43\% are expected to possess FWHM=2200-5400 \kms~and, therefore, the density of  
AGNs/QSOs is effectively cut down to $\sim$0.23 deg$^{-2}$. In any case, contaminating AGNs/QSOs  
can be further isolated through the lack of short timescale variability and 
mid-IR colour cuts \citep{mateos12, stern12, stern15}. 
Finally, ultracompact AM CVn binaries \citep{solheim10}, despite the lack of \ha~lines, possess strong broad 
\ion{He}{i} $\lambda$6678 emission which results in fake EWs through equation~\ref{eq:pew}. 
Fortunately, AM CVns are invariably placed in the \ha~colour diagram under the focus point (i.e. EW$<$0) 
as proved by synthetic colours obtained from seven Sloan AM CVn stars. 

A final word of caution must be said on our photometric system. Equation~\ref{eq:pw} has been calibrated 
assuming simulated double-peaked profiles which adequately describe quiescent BH lines (Appendix~\ref{ap:profiles}). If 
real emission lines have different shapes then FWHM$_{\rm ph}$ values may deviate with respect to true FWHMs 
by a few hundred \kms. In particular, we find that FWHMs can be overerestimated by $\sim$200 \kms~in a 
composite profile with  a broad base and narrow peak (i.e. typical of novalikes and magnetic CVs).  
On the other hand, in case of double-peaked emission superposed on broad absorption features 
(characteristic of some high-inclination WZ Sge stars) 
FWHMs are underestimated by a similar amount. In any case, we note that none of these  
represent a problem here because FWHMs in novalike/magnetic CVs never surpass our width cut-off while, 
for the latter case,  the bias goes in the direction of increasing the number of rejections.  

It should also be stressed that, because our photometric system is biased 
towards selecting stars with broad \ha~emission lines it will not find potential 
BHs with intermediate ($\approx$2-5 \msun) nor massive ($\gtrsim$10 \msun) companion stars. 
However, the contribution of these to the population of BH X-ray binaries is likely to be small, as 
indicated by both population synthesis simulations (e.g. \citealt{kalogera99, grudzinska15}) and 
observations: only three BHs with intermediate mass donors have been found among the sample of 
$\sim$60 XRTs \citep{corral16} while two BHs with masssive companions (Cyg X-1 and 
MWC 656, \citealt{casares14}) are known in the Galaxy.
 
\section{HAWKs: a Survey to discover Hibernating Black Holes} 
\label{sec:survey}

Thus far, we have shown that quiescent BH XRBs can be efficiently selected using photometric techniques, provided 
that a cut-off at FWHM$\ge$2200 \kms~is set for the width of the \ha~emission. 
In what follows, we describe a survey strategy specifically designed to discover new quiescent BHs by exploiting 
the photometric system hitherto discussed. 

\subsection{Scientific requirements and survey strategy}
\label{sec:requirements}

The so-called {\it HAlpha-Width Kilo-deg} Survey (HAWKs hereafter) hinges upon two scientific requirements.  
Firstly, we aim to measure FWHM$_{\rm ph}$ at $\approx$10\% accuracy
to match the intrinsic FWHM variability observed in quiescent BH XRTs (see Paper I). 
And secondly, the effective width of the \han~filter 
(which sets the spectral resolution of the photometric system) should be lower than our width cut-off. 
In order to ensure a clean identification of candidates we define 
the width of the \han~filter to be 1700 \kms~or 37 \AA. 
This is motivated by the fact that the bulk of contaminants 
are narrow \ha~emitters that will cluster at the \han~width 
limit in the \ha~colour diagram. Therefore, given a 10\% precision in   
FWHM measurements, our choice of \han~width implies that the great 
majority of contaminants will be placed under the 2200 \kms~cut-off with 3-$\sigma$ 
confidence and, thus, can be easily rejected.

\begin{figure}
	\includegraphics[angle=-90,width=\columnwidth]{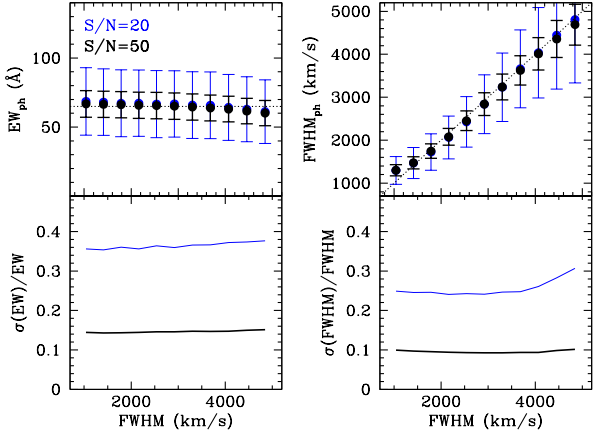}
    \caption{Photometric EWs (left) and FWHMs (right) as measured from Monte-Carlo simulations of 10$^4$ 
    synthetic spectra with added noise. The simulated spectra have EW=65 \AA~and FWHM in the range 
    1000-5000 \kms. Two examples are shown for S/N=20 (thin blue line) and S/N=50 (thick black line), as measured 
    by our three photometric filters. Filled circles and errorbars indicate the mean and $\pm$1-$\sigma$ confidence level 
    in the PDF distributions. The bottom panels  display the fractional error in EW (left) and FWHM (right) for the two 
    S/N cases. 
     } 
    \label{fig:fig10}
\end{figure}

The two previous requirements put strong constraints on photometric accuracy. 
In order to estimate the signal-to-noise (S/N) required to achieve a 10\% fractional error in FWHM  with the 37 \AA~\han~filter we decided to perform a Monte-Carlo analysis with 10$^4$ trials. Poissonian noise was injected to a grid of synthetic BH spectra, with EWs in the range 10-200 \AA~and FWHMs between 1000-5000 \kms. Simulated fluxes were then computed by convolution with the filter's response and 
used as inputs to equations~\ref{eq:pew} and \ref{eq:pw} to produce the PDFs of EW$_{\rm ph}$ and 
FWHM$_{\rm ph}$. Since the outcome is totally dominated by the narrowband filter, \han, we 
impose the additional condition that the S/N on the two other filters must be equal to the one for \han. 
It should be noted that relaxing this demand 
does neither lead to a significant improvement in S/N nor survey speed.   
Figure~\ref{fig:fig10} presents the results of the Monte-Carlo simulation for the case of an \ha~line with 
EW=65 \AA~and S/N=20 and 50. The choice of EW is motivated by the mean in the distribution of EWs 
for the current sample of dynamical BHs.  
The figure shows that S/N$\ge$50 is needed to measure FWHMs to better than 10\% 
for this specific EW. Obviously, the S/N limit increases at lower EWs (for example, S/N$\ge$60 for 
$EW\leq$50 \AA), but in what follows, we take this result as representative of the BH population 
and propose S/N=50 as the goal of the HAWKs survey. 

Quiescent BH XRTs are known to exhibit significant flickering, both in the continuum and \ha~flux, 
which can be an issue of concern \citep{hynes02, zurita03, shahbaz04}. Because flickering displays on 
characteristic time scales of $\approx$min it can be averaged out if photometric observations are split 
into short individual blocks. For instance, single S/N=50 exposures can be divided into 10 r/\hab/\han~cycles 
of S/N$\sim$16 per filter, thereby minimizing the impact of flickering variability on FWHM determinations. 
This strategy has the additional advantage of extending the dynamic range, 
avoiding the saturation of relatively bright stars.
   
HAWKs will focus on the Galactic Plane because 90\% of the $\sim$60 BH transients detected so far 
lie in the disc at $\textbar b \textbar <10^{\circ}$ \citep{corral16}. 
The survey should preferentially target selected sky fields with low 
IS extinction to maximize the surveyed volume.  
Inspection of radial extinction profiles from 2MASS \citep{marshall06} and IPHAS \citep{sale14} 
reveal some low extinction windows close to the Galactic Plane. For example, sightlines along 
$l\simeq55-75^{\circ}$ above the Plane probe the inter-arm region 
between the Perseus and Sagittarius arms to considerable depths. This area will be hereafter refereed to as 
{\it Window 1} (W1). Other 
expedient sightlines are identified below the Galactic Plane at $l\simeq40-60^{\circ}$ (W2), 
$l\simeq85-110^{\circ}$ (W3) and $l\simeq115-140^{\circ}$ (W4) 
(similar low extinction windows are reachable in the southern Galactic Plane).  
We find that these sightlines typically have $A_{r}\lesssim2.4$ up to $\approx$6 kpc (0.4 mag/kpc), even at 
low latitudes $b\sim2^{\circ}$ (see Figure~\ref{fig:fig11}). 

A dedicated survey of these fields would produce a catalogue of hibernating BH candidates with broad 
FWHM$>$2200 \kms~\ha~emission. 
In Section~\ref{sec:colour} we 
showed that significant contamination is expected from AGNs and high inclination CVs with $P_{\rm orb}<2.1$ h. 
Therefore, intensive photometric monitoring of candidates will need to be programmed in 2 h time slots 
(for instance using a robotic facility) to detect eclipsing CVs and steady (non-variable) AGNs. 
A very efficient way to identify AGNs is also through mid-IR colours: as opposed to single stars, XRBs or CVs, 
AGNs are strong mid-IR emitters which can be efficiently selected using appropriate 
colour cuts based upon {\it WISE} or {\it Spitzer} bands  (see e.g. \citealt{mateos12, stern15}). 
Finally, the remaining BH candidates can be confirmed  through photometric mass functions (Equation~\ref{eq:pmf}), 
which requires a knowledge of their orbital periods. This    
can be secured from further photometric monitoring, using specific cadence strategies optimized 
for period detection in the range 2 h - 1 d. Here the upper bound is set by the maximum P$_{\rm orb}$ for an extreme 
15 \msun~BH (see e.g. \citealt{fryer01}, also \citealt{belczynski10}) to produce FWHM$>$2200 \kms. 
The final output of the HAWKs  survey will thus be a list of confirmed BHs with PMF$\ge$3 \msun. 

\subsection{The hibernating BH population}
\label{sec:census}

A key question to follow is how many quiescent BHs can be discovered by HAWKs and this depends primarily 
on the size of the hibernating population. To address this issue we first look at empirical estimates. 
Modern extrapolations of the number of X-ray transients detected thus far suggest a Galactic population of 
$\approx2000$ dormant BHs \citep{romani98,corral16}. 
These estimates, however, are most likely biased low because of survey incompleteness and a number of complex selection 
effects. For example, arguments have been presented for the existence of  
a hidden population of BHs with cold globally stable discs and, thus,  
suppressed outburst activity  
\citep{menou99}. Furthermore, it has been shown that BH XRTs with short 
orbital periods $\lesssim$4 h 
may be concealed because of low peak X-ray outburst luminosities \citep{wu10, knevitt14}.  
Also, the paucity of dynamical BHs with orbital inclinations  $>75^{\circ}$ strongly 
indicates that high inclination BHs are X-ray obscured by their accretion discs and thus difficult to 
detect (\citealt{narayan05}, see also \citealt{corral13}). 
Considering this we believe the observational projections are underestimated by a factor of a few.   

Standard binary population models, on the other hand,  
have problems reproducing the empirical numbers because of the energetics of the Common Envelope (CE) 
phase. It is exceedingly difficult for the low-mass companion to eject the envelope of the massive star  
and most progenitor binaries are predicted 
to end up in mergers (see e.g. \citealt{portegies97}). 
Other simulations involving less standard CE parameters or alternative formation paths do, however, predict 
$10^3-10^4$ BHs in the Galaxy, in better agreement with observations (see review in \citealt{li15}). 
Given the above, we here decide to adopt a Galactic population of 5000 hibernating BHs. 

 \subsection{Estimated number of BHs to be revealed by HAWKs}
\label{sec:number}

Armed with the census of hibernating BHs we can now provide 
an estimate of the number to be discovered by HAWKs. Here we assume 
 that the density of BHs follows an exponential distribution 
 
 \begin{equation}
 \rho=\rho_0 \exp(-z/h)
   \label{eq:density}
 \end{equation}
 
 \noindent
 with $z$ being the distance perpendicular to the Galactic Plane and $h$ the scale height. We start by adopting 
 $h\sim0.167$ kpc, consistent with the scale height of stars in the thin disc \citep{binney08}. 
 The normalization for a population of 5000 BHs thus implies a Galactic mid-plane density 
 $\rho_0=24$ kpc$^{-3}$. A crucial aspect in 
 this calculation is the radial distribution of interstellar extinction along our proposed sightlines. 
 As an example, we have examined IPHAS extinction profiles for sky region W1. 
  
 \begin{figure}
	\includegraphics[angle=-90,width=\columnwidth]{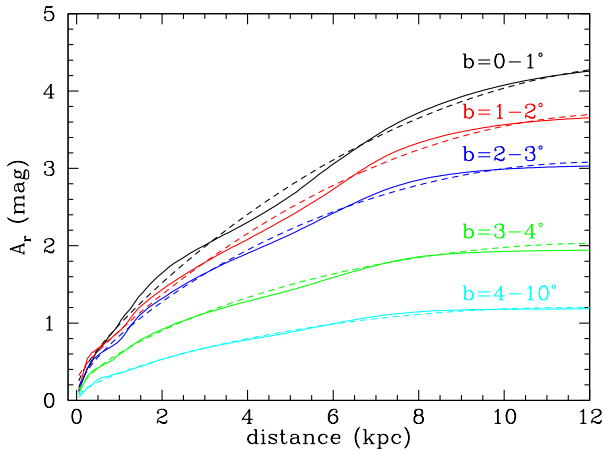}
    \caption{Average radial extinction profiles for window W1 ($l=55-75^{\circ}$) and five latitude bands. Dashed lines 
    indicate the best fitting curves to each profile. 
     } 
    \label{fig:fig11}
\end{figure}

 Radial profiles were downloaded from \citet{sale14} (http://www.iphas.org/extinction/) 
 for six latitude bands between $b=0^{\circ}$ and 5$^{\circ}$ and these have been averaged  
 within the longitude interval of W1 to produce a mean extinction profile per latitude band. 
 IPHAS provides $A_0$, the monochromatic extinction at 5495 \AA, and, therefore, these values were converted 
 into r-band extinction through $A_{r}=0.748~A_0$ \citep{cardelli89}.   
 We find that the extinction profiles can be modeled with  
 a combination of a quadratic and power functions 
following $A_{r} (d) =a_{0}~d \times (1+a_{1}~d) + a_{2}~d^{a_{3}}$, with $d$ being the  
 distance along a given sightline and $a_{0}$-$a_{3}$ the fitting coefficients. 
 Figure~\ref{fig:fig11} displays the average radial extinction 
 profiles together with the best model fits for our five representative latitude slices. 
 Here, because IPHAS extinction maps are constrained to $\textbar b \textbar <5^{\circ}$, we have taken the 
 conservative approach of extending the profile of slice $b=4^{\circ}-5^{\circ}$ 
 to the entire latitude strip $b=4^{\circ}-10^{\circ}$.
 
 The maximum surveyed distance for a given sightline is provided by the distance modulus equation 
 
 \begin{equation}
\left(\frac{d}{\rm kpc}\right)=10^{~\left[0.2~\left(m_{r}-M_{r}-A_{r} (d)~\right)-2\right]}
   \label{eq:distance}
 \end{equation}
 
 \noindent
 where we adopt the absolute magnitude of the prototypical binary A0620-00 ($M_r=6$) as representative of 
 the BH XRT population. The total number of BHs will thus be obtained by integrating  
equation~\ref{eq:density} over window W1, subdivided into five 
 latitude bands (i.e.  $b=0^{\circ}$-1$^{\circ}$, $1^{\circ}$-$2^{\circ}$, $2^{\circ}$-$3^{\circ}$, $3^{\circ}$-$4^{\circ}$ 
 and $4^{\circ}$-$10^{\circ}$), and to a depth given by equation~\ref{eq:distance}.  Table~\ref{tab:tab2} 
 lists the total number of BHs in field W1 for three different survey magnitudes. It also provides a breakdown of the number of BHs per latitude slice.  
 
 \begin{table}
	\centering
	\caption{Maximum probed distance and number of BHs in window W1 for scale height $h=0.167$ kpc and three survey magnitudes}
	\label{tab:tab2}
	\begin{tabular}{ccccccc}
		\hline
		b &  \multicolumn{2}{c}{r=21} & \multicolumn{2}{c}{r=22} & \multicolumn{2}{c}{r=23} \\
 		   & d & Number  & d & Number & d & Number \\ 
                   & (kpc)  & BHs &  (kpc)  & BHs & (kpc) & BHs \\
		\hline
$0^{\circ}-1^{\circ}$   & 3.6  & 2.0  & 4.7  & 4.3  & 6.0 & 8.5 \\ 
$1^{\circ}-2^{\circ}$   & 3.8  & 1.7  & 5.0  & 3.5 & 6.6 & 6.7 \\
$2^{\circ}-3^{\circ}$   & 4.1  & 1.6  & 5.5  & 2.9 & 7.4 & 5.1 \\
$3^{\circ}-4^{\circ}$   & 5.0  & 1.7  & 7.1  & 2.9 &10.2 & 4.4 \\
$4^{\circ}-10^{\circ}$ & 6.3  & 4.6  & 9.4  & 6.0 & 15.3 & 6.6 \\
\hline
TOTAL BHs& - & 12 & - & 20 & - & 31 \\
\hline
	\end{tabular}
\end{table}

   It should be noted, however, that the numbers given in Table~\ref{tab:tab2} are likely to be conservative 
   for several reasons. First, 
    we have assumed that BH XRTs follow the scale height of the thin disc whilst recent studies support a 
   larger value  $h\simeq0.69$ kpc, consistent with the presence of significant BH kick velocities 
   \citep{repetto17}.  
   Remarkably,   while the scale height does not affect the 
   number of BHs at r=21, it does have a significant impact at higher depths, with a 30\% increase 
   for r=22 (and a factor two for r=23). This is reflected in Table~\ref{tab:tab3} where we list the number of BHs 
   for $h=0.69$ kpc and three survey magnitudes\footnote{A mid-plane density $\rho_{0}=6$ Kpc$^{-3}$ 
   has been adopted here so that the total number of BHs is 5000.}.
   In this regard, 
   it is interesting to note that a direct comparison of the number of BHs per latitude band with binary population 
  simulations can readily demonstrate the existence of high natal kicks during BH formation 
  \citep{jonker04,repetto12, repetto17}. As we see in Tables~\ref{tab:tab2} and \ref{tab:tab3}, this is achievable 
  with a single 200 sqr deg field, such as W1, down to a depth r$\sim22$. 

 \begin{table}
	\centering
	\caption{Number of BHs in window W1 for scale height $h=0.69$ kpc and three survey magnitudes}
	\label{tab:tab3}
	\begin{tabular}{cccc}
		\hline
		b &  r=21 &  r=22 & r=23 \\ 
 		   & Number & Number &  Number \\ 
                   & BHs  & BHs & BHs \\
		\hline
$0^{\circ}-1^{\circ}$   &  0.6  & 1.2  & 2.5 \\ 
$1^{\circ}-2^{\circ}$   &  0.6  & 1.3  & 2.9 \\
$2^{\circ}-3^{\circ}$   &  0.7  & 1.6  & 3.5 \\
$3^{\circ}-4^{\circ}$   &  1.1  & 2.8  & 6.7 \\
$4^{\circ}-10^{\circ}$ &  8.3  & 19.0 & 44.3 \\
\hline
TOTAL BHs& 11 & 26 &  60 \\
\hline
	\end{tabular}
\end{table}
  
And secondly, 
   because of the lack of accurate extinction profiles for latitudes $b>5^{\circ}$ we have extended the IPHAS 
  profile for $b=4-5^{\circ}$ to $b=4^{\circ}-10^{\circ}$. If we take instead a typical (but less accurate) NIR extinction 
  profile for ($l=65^{\circ}$, $b=7.5^{\circ}$) as given in \citet{marshall06} and convert it into r-band extinction through 
  $A_{r}=6.68 A_{\rm K}$ \citep{cardelli89}, we find that the number of BHs is further boosted 
(e.g. $\approx$41 for r=22). 
  On the other hand, we should bear in mind that, as shown in Section ~\ref{sec:cutoff}, our 
  $FWHM\ge2200$ \kms~cut-off will select $\sim$46\% of the BH population contained in a 
  given survey volume and, therefore,  HAWKs would only be able to detect half of the numbers 
listed in Table~\ref{tab:tab3}. 
     
   To conclude this Section we look at the number of outliers expected for window W1. In the case of CVs 
  we assume an exponential density profile in the vertical direction with $\rho_0=2\times10^4$ kpc$^{-3}$ 
  \citep{politano96, pretorius12, burenin16} and $h=0.26$ kpc, appropriate for the disc scale height of old, short 
  period CVs \citep{pretorius07}.  Following \citet{warner87} we adopt an absolute magnitude $M_{\rm V}$=9, 
  characteristic of  short period CVs, and a typical quiescent colour $(V-R)\simeq0$. Integrating 
  equation~\ref{eq:density} over window W1 yields 4600 CVs for a survey depth r=22. But, since only 0.1\% will   
  possess FWHM>2200 \kms~(Section~\ref{sec:cutoff}) we expect to select $\approx$5 CVs.  
  Regarding AGNs, in Section~\ref{sec:contaminants} we predicted $\approx$0.23 per sqr deg for V$<$22 which 
  results in $\approx$46 AGNs for window W1. We note, however, that this is an upper limit because it neglects 
  Galactic absorption which can amount to several magnitudes in the Plane.  
 Adopting the extinction profiles derived in Figure~\ref{fig:fig11} for W1 and using statistics 
 from the Veron Catalogue we estimate that our sample will contain $\approx$38 AGNs. 
 
  In summary, for a  survey magnitude r=22 and a BH scale height 
  $h\sim0.69$ kpc we predict HAWKs will select $\approx$56 candidates with FWHM>2200 \kms~in 200 sqr deg. 
  These will include $\approx$5 CVs, $\approx$38 AGNs and $\approx$13 hibernating BHs which can be singled out 
  through photometric variability. 
  Contaminating AGNs can also be detected and weed out through mid-IR colours. 
  A survey of 800 sqr deg would therefore produce $\approx$50 new 
  dynamical BHs i.e. a three-fold improvement over the currently known sample. 
  
\section{Discussion}
\label{sec:discussion}

BH XRTs are presently discovered at a rate of $\sim$1.7 yr$^{-1}$ but less than a third remain sufficiently bright 
in quiescence for radial velocity analysis. As a result, after 50 years of careful 
scrutiny of the X-ray sky only 17 BHs have been dynamically confirmed and, therefore, about a century seems 
necessary to simply double the number.  A prime goal in the field would thus be to unveil 
a large homogeneous sample of secure (dynamical) quiescent BHs. 
To this end we have presented 
a novel approach to uncover an unprecedented large sample of dormant BHs. 
This can be achieved through a deep ($r\sim$22) survey using a combination of carefully chosen H$\alpha$ filters. 
The filters are specially customized to resolve H$\alpha$ widths FWHM$>$2200 \kms, a limit which, as we showed in 
Section~\ref{sec:contaminants}, would allow selecting $\sim$46\% of the hidden BH population while filtering out the 
vast majority of contaminants. 
Furthermore, the combination of \ha~widths with orbital information (supplemented by    
photometric follow-up of candidates), will lead to {\it photometric} mass functions and thus dynamical confirmation 
of hibernating BHs.  

The new concept presented in this paper (i.e. photometric BH selection) opens the door to 
weigh BHs through imaging and, thus, search for BHs in large volumes to unprecedented depths, 
a potential breakthrough in the field. 
We estimate that $\approx$13 BHs can be unveiled in an area of 200 sqr deg and, therefore, a survey organized 
around this technique can lead to an order of magnitude improvement on dynamical BH statistics in just a few years. 
This would allow constraining, not only the space properties of the population (i.e. Galactic Plane density, scale height 
and total number) but also the distribution of orbital 
periods and, ultimately, the BH mass spectrum. 
These three observables encode key information on BH formation and evolution. 
For example, the total number of BHs is very sensitive to binary evolution parameters 
such as the efficiency of common envelope ejection, mass-loss rate of He stars or the 
survival rate after supernova explosion (e.g. \citealt{kalogera99,podsiadlowski03,kiel06}). 
Meanwhile, the observed scale height is driven by the presence of natal kicks, a current hot topic in the field 
\citep{jonker04,fragos09,repetto17, mirabel17, casares17}. 
It should be stressed here that, even if the total number of discovered BHs turns out to be lower than predicted, 
HAWKs will still set stringent constraints on BH formation models. 

On the other hand, the BH mass spectrum strongly depends on poorly constrained supernova physics, 
starting with the nature of the explosion mechanism itself \citep{fryer12,ugliano12, kochanek14}.  
An expanded sample will, for instance, allow quantification of the significance of the gap in the mass distribution
\citep{ozel10, farr11,kreidberg12} or test the claimed BH mass-period relation \citep{lee02}.
The latter was proposed to stem from a correlation with BH spin and thus, if true, it should be closely 
connected to the production of long-duration gamma-ray bursts. 
It should also be noted that, because HAWKs is biased towards \ha~emitters, BH X-ray binares with hotter 
early-type companions  will be missed. 
However, as mentioned earlier, both population synthesis models and current observations  
strongly suggest that the impact of the latter on the distribution of BH masses is likely to be negligible. 

Finally, the orbital period distribution is shaped by the length of time BH X-ray binaries spend in a given 
period bin and, therefore, on the relative importance of gravitational radiation versus magnetic braking 
\citep{knigge11}, with perhaps other unknown angular momentum loss mechanisms also playing a role, 
as suggested by recent observations of fast orbital period decays \citep{gonzalez14}. 
As an example, the elusive minimum period spike predicted by CV evolution models was a subject of debate for 
decades until a faint population of quiescent CVs was revealed by SLOAN \citep{gansicke09}.  
Similarly, the outcome of the HAWKs survey will allow for constraints on the lower limit in the hibernating BH period 
distribution, searching for evidence of a gap and a possible accumulation of systems around a minimum period.   

\begin{figure}
	\includegraphics[angle=0,width=\columnwidth]{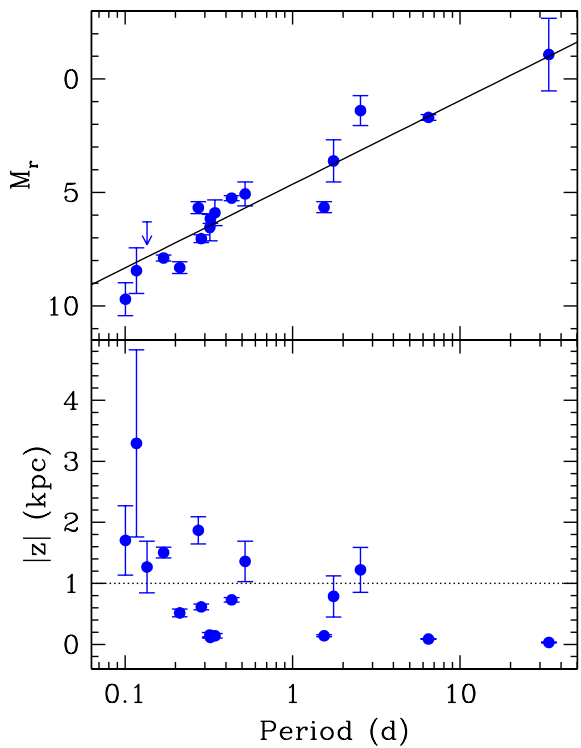}
    \caption{Top: Absolute magnitudes of quiescent BH XRTs with low-mass companion stars as a function of 
    orbital period. The best least squares fit is overplotted. Bottom: Vertical elevation above the Galactic plane 
    versus orbital period. The dotted horizontal line marks the limit of the Galactic thin disc.  
     } 
    \label{fig:fig12}
\end{figure}

The study of BH XRTs with short orbital periods has strategic interest because these might represent 
a large fraction of the whole hibernating population, with different bulk properties than the existing sample. 
To illustrate this, we have compiled in Table~\ref{tab:tab4} quiescent properties of the known 
BH XRTs with low-mass companion stars. Two binaries (GRS1915+105 and GX 339-4) have not yet returned 
to quiescence after discovery and, thus, we decided to adopt their deepest optical magnitudes  
reported to date, which correspond to epochs of minimum X-ray activity. Because of their long orbital 
periods and luminous companion stars, we expect irradiation effects to be negligible at these low levels. 
Therefore, we tentatively 
adopt these magnitudes as representative of their true quiescent luminosity. In any case, it should be noted that 
the error in absolute magnitude for these two sources  
is dominated by uncertainties in the distance and, in the case of GRS1915+105, also reddening. 

Figure ~\ref{fig:fig12} displays the evolution of absolute magnitude (M$_{r}$) and elevation 
above the Galactic plane ($z$) as a function of orbital period. 
Interestingly, we observe that M$_{r}$ is tightly 
correlated with orbital period across 10 magnitudes 
(i.e. 4 decades in luminosity), with a correlation coefficient -0.95. A least squares fit yields   

 \begin{equation}
M_{r}= 4.64(0.10)-3.69(0.16) \log P_{\rm orb}~(d), 
   \label{eq:magnitude}
 \end{equation}

 \noindent
 a relation which, in turn, may prove useful to infer distances to BH XRTs when the orbital period is known. 
 The mean absolute magnitude for the period interval 2-4 h is $M_{r}\simeq9$, i.e. 3 magnitudes fainter than that of  
 BH XRTs at the peak of the period distribution. This makes short period BHs difficult to detect by 
 HAWKs unless the search is pushed to deeper limits $r\sim$24. On the other hand, 
 the bottom panel in Figure ~\ref{fig:fig12} shows that short period BHs tend to be located in the Galactic thick disk 
 at $z\ge$1 kpc and, consequently, are less affected by interstellar absorption. Although this may be a 
 selection effect (short period binaries might be X-ray dimmer 
 and, therefore, more heavily obscured in the plane) it implies that short period BHs are 
 best targeted at higher Galactic latitudes. 
At S/N$\sim$15 for r$\sim$24 HAWKs should still be able to identify short period BHs, given their   
typically large FWHM$\approx$3000-4000 \kms~(see Figure~\ref{fig:fig9}), or, at least, put 
useful constraints on their abundance.  

It should also be mentioned that HAWKs is sensitive to the discovery of massive (high inclination) 
NSs in XRBs and Redbacks in the disc state, the evolutionary link between Low Mass X-ray Binaries and 
recycled millisecond pulsars \citep{roberts13}. The construction of the mass distribution of NSs is one of the fundamental 
experiments in high energy astrophysics from which new light on the composition of ultradense matter can 
be shed \citep{ozel16}. In addition, as shown by Figure~\ref{fig:fig8}, the \ha~colour diagram is very efficient at 
discriminating \ha~emitting objects from normal stars. The survey will thus render a flux-limited sample of 
other populations of \ha~emitters down to unprecedented depths which will allow a plethora of ancillary studies. 
In particular it will greatly expand the number of WZ Sge stars with secure periods below the gap, a key 
sample to constrain formation and evolutionary channels in CVs. This sample will likely contain massive 
accreting white dwarfs, which will make excellent candidates for Type Ia supernova precursors 
\citep{littlefair08, maoz14}.  
As a matter of fact, new generation synoptic surveys  (e.g. the {\it Large Synoptic Survey Telescope} 
{\it LSST}, \citealt{abell09}) and wide-field multi-object spectrographs (e.g. WEAVE, \citealt{dalton12}) 
will secure optical colours, variability information and classification spectra of 
interesting subsets of \ha~emitters selected from HAWKs. For example routine light curves and spectra of sources with 
FWHM$_{\rm ph}>1800$ \kms~will likely reveal large numbers of eclipsing white dwarfs and NS binaries, 
(ideal for accurate mass determinations) plus many more BHs, a factor $\sim$1.8 larger than hitherto quoted.

\section{Conclusions}
\label{sec:conclusions}

We have presented HAWKs, a photometric survey with custom \ha~filters designed for the efficient detection 
of dormant BHs in large fields of view to unprecedented depth. The main conclusions of the paper are as follows:

\begin{itemize}

\item{} FWHMs of \ha~emitting stars can be recovered from photometric observations using a combination of  
r-band continuum filter and two \ha~filters (one broad and another narrow), all centered at the rest wavelength of 
\ha. We define a photometric system based on these three filters and show how \ha~emitters can be effectively 
separated from normal stars in a colour-colour diagram, independently of their SEDs and interstellar reddening. 

\item{} We find that a width cut-off at FWHM$>$2200 \kms~allows the selection of $\sim$46\% of hibernating BHs while 
filtering out the great majority of \ha~emitters. Only AGNs at particular redshift bands and a small number of 
eclipsing short period CVs can contaminate the sample. Fortunately, these are easily flagged from variability 
properties in 2 h light curves, with AGNs being steady sources while CVs display deep eclipses. 
AGNs can also be identified from mid-IR colour cuts.  

\item{} BH candidates selected by our photometric system will be dynamically confirmed through {\it Photometric 
Mass Functions}. These are obtained by combining the photometric FWHMs with orbital period information derived 
through photometric variability. This strategy establishes the basis for HAWKs, an {\it HAlpha-Width Kilo-deg} 
Survey of the Galactic plane. 

\item{} We predict HAWKs will discover $\approx$13 hibernating BHs in a survey of 200 sqr deg at depth r=22. 
This calculation assumes a Galactic population of $\sim$5000 dormant BHs, distributed with a scale height 
h=0.69 kpc and similar properties as the currently known sample i.e. same distribution of quiescent 
luminosities, orbital periods and BH masses. Based on these numbers we expect an order of magnitude 
improvement in the statistics of dynamical BHs through a survey area of $\approx$1 kilo sqr deg, a goal which is 
achievable in only a few years.

\item{} The previous estimates are, however, subject to considerable uncertainty due to possible unknown 
biases affecting the existing sample. In any event, only through a deep specialized 
survey such as HAWKs is it possible to constrain the size and properties of the hidden BH population. 
For example, both the scale height and population size  
are very sensitive to the number of BHs detected at Galactic latitudes $b\sim5^{\circ}-10^{\circ}$ and depths 
r$\sim22-23$. These parameters could be constrained from a limited survey area of $\sim$200 sqr deg, such as the one 
proposed for W1. We believe the outcome of HAWKs may revolutionize the study of stellar-mass BHs by significantly 
boosting the census, setting stringent constraints on the properties of the population and paving the way for 
demographic studies.
\end{itemize}

\section*{Acknowledgements}

This work was supported by the Leverhulme Trust through the Visiting Professorship Grant VP2-2015-046.  
Also by the Spanish Ministry of Economy, Industry and Competitiveness (MINECO) under the 2015 Severo Ochoa Program MINECO SEV-2015-0548. 
We would like to acknowledge the hospitality of the Department of Physics of the 
University of Oxford, where this work was performed during a sabbatical visit. The author is also grateful to Mansfield College for the support provided by a Visiting Fellowship and the kindness of its staff. 
We thank P.A. Charles, R. Fender, T. Maccarone, T. Mu\~noz-Darias, M.A.P. Torres, P. Ghandi and J.A. Fern\'andez-Ontiveros for many interesting discussions and comments to the manuscript. 
We also thank the anonymous referee for helpful suggestions that have improved the quality of the paper.

\begin{landscape}
\begin{table}
	\centering
	\caption{Population fraction selected as a function of FWHM cut-off values.}
	\label{tab:tab1}
	\begin{tabular}{cccccccc}
		\hline
		Model & FWHM & FWHM &FWHM & FWHM & FWHM & FWHM & FWHM \\ 
            Population  &  $>1000$ \kms  & $>1500$ \kms  & $>1800$ \kms & $>2000$ \kms  & $>2200$ \kms & $>2400$ \kms & $>2600$ \kms \\
		\hline
CV-1 & 0.931 & 0.505 & 0.136 & 0.016 & $3\times 10^{-4}$ & 0 & 0 \\ 
CV-2 & 0.996 & 0.769 & 0.224 & 0.038 & 0.001 & $1\times 10^{-4}$& $4\times 10^{-5}$ \\
BH & 0.999 & 0.965 & 0.811 & 0.633 & 0.459 & 0.327 & 0.238 \\ 
\hline
	\end{tabular}
\end{table}
\end{landscape}

\begin{landscape}
 \begin{table}
	\centering
	\caption{Quiescent properties of BH XRTs with low mass companion stars}
	\label{tab:tab4}
	\begin{tabular}{lccccccc}
			\hline
Source & P$_{\rm orb}$ & d &  m$_{\rm r}$ & E(B-V) & M$_{\rm r}$ & b & $\textbar z \textbar=d~\sin \textbar b \textbar$ \\ 
 		           & (days)            & (kpc) &  (mag) &          & (mag)        & (deg) & (kpc) \\ 
 		\hline
GRS 1915+105   &   33.85   &   9.0$\pm$2.0      & 28.3 &  6.3$\pm$0.5 & -1.08$\pm$1.6 &-0.2 & 0.03$\pm$0.01 \\
V404 Cyg           &  6.4713  &   2.39$\pm$0.14  & 16.6  & 1.3   & 1.7$\pm$0.1 & -2.1&  0.09$\pm$0.01 \\
BW Cir               &  2.5445   &          18-33         &  20.7  & 1.0   &  1.4$\pm$0.7 & -2.8  & 1.22$\pm$0.37 \\
GX 339-4           &  1.7557   &            6-15         &  21.5  & 1.2   & 3.6$\pm$0.9 & -4.3 & 0.79$\pm$0.34 \\
XTE J1550-564  &  1.5420  &   4.5$\pm$0.5     &  22     & 1.33  & 5.7$\pm$0.2 & -1.8 & 0.14$\pm$0.02 \\
N. Oph 77           &  0.5213  &   8.6$\pm$2.1     & 20.9   &  0.5  &  5.1$\pm$0.5 & 9.1 & 1.36$\pm$0.33 \\
N. Mus 91           &  0.4326  &   5.9$\pm$0.3     & 19.8   & 0.3   & 5.3$\pm$0.1  &-7.1 & 0.73$\pm$0.04 \\
GS 2000+25       &  0.3441  &   2.7$\pm$0.7     & 21.3   & 1.4   & 5.9$\pm$0.6  & -3.0 & 0.14$\pm$0.04 \\
A0620-00            &  0.3230  &  1.06$\pm$0.10 & 17.1    & 0.35  & 6.2$\pm$0.2 & -6.2  & 0.11$\pm$0.01 \\
XTE J1650-500   &  0.3205  &  2.6$\pm$0.7     & 22.1    & 1.5   & 6.6$\pm$0.6  & -3.4 & 0.15$\pm$0.04 \\
N. Vel 93              &  0.2852  &  3.8$\pm$0.3    &  20.4    & 0.2   &  7.0$\pm$0.2 & 9.3 & 0.61$\pm$0.05 \\
XTE J1859+226   &  0.2740  & 12.5$\pm$1.5   &  22.5    & 0.58  & 5.7$\pm$0.3 & 8.6 & 1.87$\pm$0.22 \\
GRO J0422+320  &  0.2122  &  2.5$\pm$0.3    &  21.0    & 0.3   & 8.3$\pm$0.3 & -11.9 & 0.52$\pm$0.06 \\
\hline
XTE J1118+480    & 0.1699  &  1.7$\pm$0.1     & 19.1   &  0.024 & 7.9$\pm$0.1 & 62.3 & 1.51$\pm$0.09 \\
Swift J1753-127    & 0.1352  &   6.0$\pm$2.0   & $>$21 & 0.34    & $>$6.3          & 12.2 & 1.27$\pm$0.42 \\
 Swift J1357-0933 & 0.1170   &      2.3-6.3       & 21.7   & 0.037  &  8.5$\pm$1.0 & 50.0 & 3.29$\pm$1.53 \\
 MAXI J1659-152  & 0.1006   &  6.0$\pm$2.0    & 24.2   & 0.26    & 9.7$\pm$0.7  & 16.5 & 1.70$\pm$0.57 \\
\hline
	\end{tabular}
\end{table}
{\bf Notes to table:} M$_{\rm r}$ is obtained from the distance modulus equation, assuming average Galactic reddening 
$R_{\rm V}=A_{\rm V}/E(B-V)=3.1$ and $A_{\rm r}=0.748~A_{\rm V}$ \citep{cardelli89}. 
Remaining values are taken from the BlackCAT catalogue \citep{corral16} except for: 

\begin{itemize}
\item{}{\bf GS 1915+105:} m$_{\rm r}$ is derived from m$_{\rm K}$=13.1 (minimum X-ray activity; \citealt{neil07}), 
assuming a de-reddened colour (m$_{\rm r}$-m$_{\rm K}$)$_0$=2.47, appropriate for a K5 III \citep{johnson66,ziolkowski17}, $A_{K}/A_{r}=0.15$ and E(B-V) from \citet{chapuis04}. 

\item{}{\bf BW Cir:} $d$ from \citet{casares09}, assuming a 30\% uncertainty. 

\item{}{\bf GX 339-4:} $d$ from \citet{hynes04} and m$_{\rm r}$ from a reported episode of minimum X-ray activity \citep{lewis12}.

\item{}{\bf Swift J1753-127:} $d$ from \citet{cadolle07}. 

\item{}{\bf Swift J1357-0933:} $d$ range constrained from a lower limit in \citet{mata15} and an upper limit in \citet{shahbaz13}. 

\item{}{\bf MAXI J1659-152:} $d$ from \citet{jonker12}, m$_{\rm r}$ from \citet{corral17}.
\end{itemize}

\end{landscape}








\appendix

\section{Synthetic double-peak profiles, fluxes and magnitudes}
\label{ap:profiles}

We have created synthetic X-ray binary spectra by adding a symmetric double-peaked \ha~profile to a linear continuum 
of slope 5$\times10^{-4}$ ergs cm$^{-2}$ s$^{-1}$ \AA$^{-1}$. This particular slope reproduces well the reddened spectra 
of X-ray binaries such as V404 Cyg, BW Cir or GS 2000+25. The continuum has been normalized to unity at the rest 
wavelength of \ha~($\lambda_{0}=6562.76$)~i.e. 

\begin{equation}
f_{\rm cont}=0.0005\times\left(\lambda-\lambda_{0}\right) + 1  
\label{eq:cont}
\end{equation}

\noindent
 where $\lambda$ is the wavelength in angstroms. The \ha~line profile is modeled as 
 
\begin{equation}
f_{H\alpha}= 0.2 \left(\frac{EW}{\sigma}\right)\times\left({\rm e}^{-\frac{\left(\lambda-\left(\lambda_{0}-DP/2\right)\right)^{2}}{2\sigma^2}} + {\rm e}^{-\frac{\left(\lambda-\left(\lambda_{0}+DP/2\right)\right)^{2}}{2\sigma^2}}\right)
 \label{eq:ha}
\end{equation}

\noindent
where $DP$ stands for the double peak separation of the \ha~profile. The area of the profile is set to be the EW of the 
\ha~line. Model profiles were computed for $DP=10-60$ \AA~and, for practical purposes, we have adopted 
$\sigma=DP/3$. 

\begin{figure}
	\includegraphics[angle=-90,width=\columnwidth]{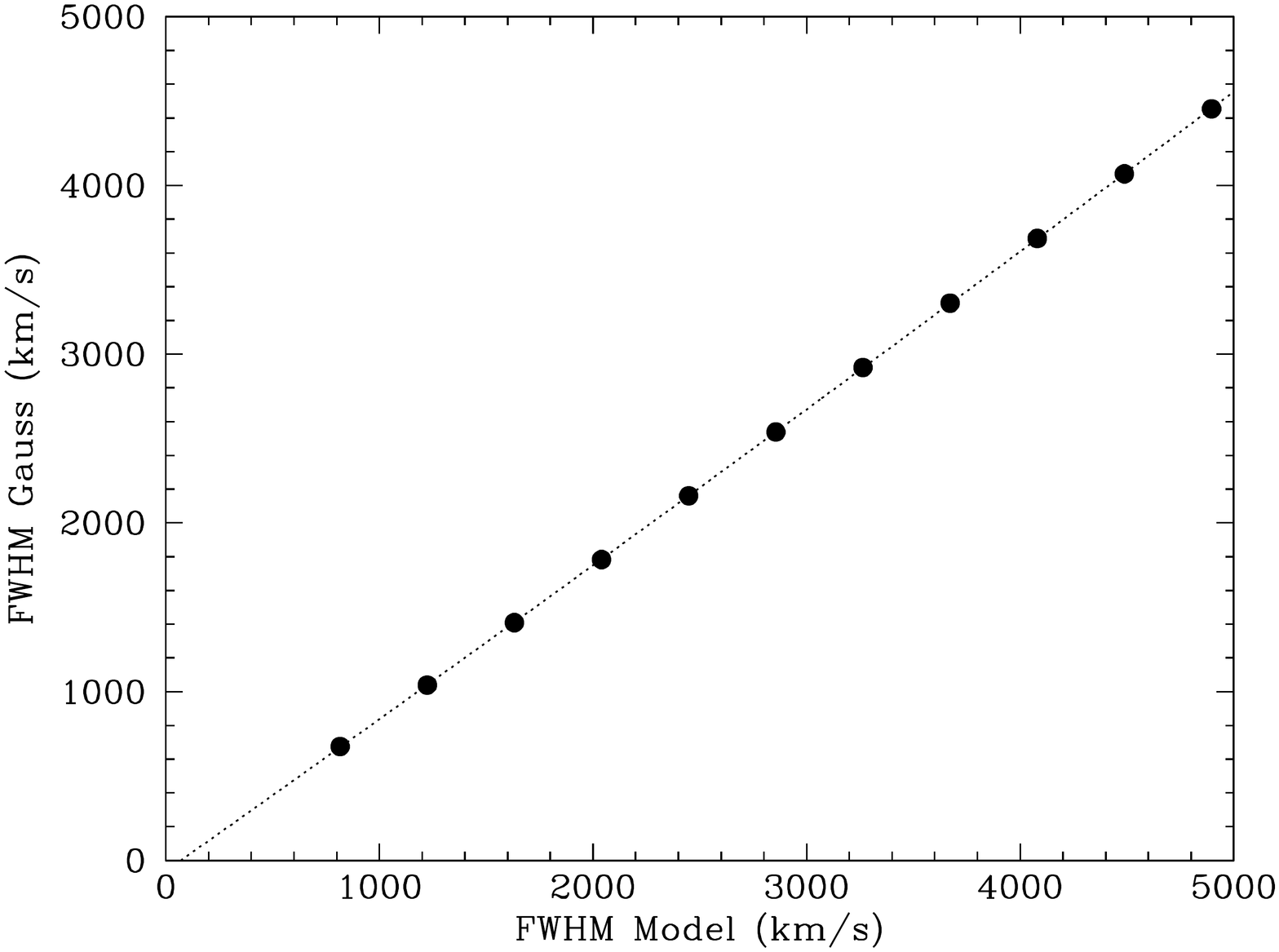}
    \caption{FWHMs calculated from a Gaussian fit to synthetic profiles versus model FWHMs. The dotted line 
    depicts the best quadratic fit.} 
    \label{fig:a1}
\end{figure}

By definition, the FWHM of the model profile (FWHM$_{m}$) is equivalent to the double peak separation plus the 
FWHM of any of the Gaussians than make up the profile i.e. $FWHM_{m}=DP+2 \sigma \sqrt{2 \ln 2}= 
1.785\times DP$. 
On the other hand, it is argued in Paper I that a single Gaussian fit provides a practical and robust determination of 
the line FWHM 
for the case of real data.  Therefore, a conversion between formal (model) FWHMs and values derived 
through Gaussian fits (FWHM$_{g}$) is required. Consequently, we have fitted 
Gaussian functions to our synthetic profiles and derived FWHM$_{g}$ values, which are listed in Table~\ref{tab:a1} 
and plotted in figure~\ref{fig:a1}. The quadratic fit 

\begin{equation}
FWHM_{g}=-62.84+0.8942~FWHM_{m} + 5.9\times10^{-6}~FWHM_{m}^2  
\label{eq:fwhm}
\end{equation}

\noindent
allows obtaining Gaussian FWHMs from formal FWHMs to within 0.6 \%.  

\begin{table}
	\centering
	\caption{Model FWHMs compared to FWHMs from Gaussian fits.}
	\label{tab:a1}
	\begin{tabular}{ccc}		\hline
		DP & Model FWHM  & FWHM Gaussian \\ 
		(\AA) & (km s$^{-1}$) & (km s$^{-1}$) \\
		\hline
10 &   816 &   675 \\ 
15 & 1224 & 1039 \\ 
20 & 1632 & 1409 \\ 
25 & 2040 & 1783 \\ 
30 & 2448 & 2161 \\ 
35 & 2856 & 2539 \\ 
40 & 3264 & 2921 \\ 
45 & 3672 & 3303 \\ 
50 & 4080 & 3686 \\ 
55 & 4488 & 4069 \\ 
60 & 4896 & 4454 \\ 
\hline
	\end{tabular}
\end{table}

The flux measured by a given filter is obtained through integrating the synthetic spectra over wavelength, after convolution 
with the transmission curve of the filter ($T_{H\alpha_b}$), e.g. in the case of filter \hab  

\begin{equation}
F ({H\alpha_b})= \int{T_{H\alpha_b}\times \left(f_{cont}+f_{H\alpha}\right) d\lambda}
 \label{eq:convolution}
\end{equation}

\noindent
Both synthetic spectra and filter response curves are previously sampled to a common bin size of 1 \AA~pix$^{-1}$. 
The fluxes so obtained are subsequently converted into instrumental magnitudes through the expression 

\begin{equation}
m_{H\alpha_b}= -2.5 \log F_{H\alpha_b}
 \label{eq:mag}
\end{equation}

\noindent
and used to construct the \ha~colour diagram presented in Section~\ref{sec:colour}.

\section{Photometric EW}
\label{ap:pew}

We assume a simplified X-ray binary spectrum consisting of a flat continuum and an \ha~emission line with square 
profile. We call 
$f_{\rm cont}$ the continuum flux density level and $A$ the integrated line flux i.e. $A=f_{\rm cont}\times EW$ where 
EW stands for the equivalent width of the \ha~line. We measure the flux of the X-ray binary using both, a broad r-band 
and an \ha~filter. These filters have equivalent widths $W_{\rm r}$ and $W_{H\alpha_b}$ respectively, with 
$W_{\rm r}>W_{H\alpha_b}>EW$. For simplicity we also assume the two filters are perfectly squared i.e. Heaviside step 
functions with maximum 100\% transmission.  The flux measured by each filter will be 

\begin{equation}
F_{\rm r}= A + f_{\rm cont}~W_{\rm r} = A \left( 1+ \frac{W_{\rm r}}{EW} \right) 
 \label{eq:fr}
\end{equation}

\noindent
and 

\begin{equation}
F_{H\alpha_b}= A \left(1+\frac{W_{H\alpha_b}}{EW} \right) 
 \label{eq:fb}
\end{equation}

\noindent
By computing the ratio between equations~\ref{eq:fb} and ~\ref{eq:fr} we manage to cancel out $A$ and obtain 

\begin{equation}
EW=\frac{W_{r} \times \left(\frac{F_{H\alpha_b}}{F_r}\right) - W_{H\alpha_b}}{1-\left(\frac{F_{H\alpha_b}}{F_r}\right)}.  
\label{eq:ew}
\end{equation}

\noindent 
Finally we define  

\begin{equation}
EW_{ph}=C_{1}~EW
\end{equation}

\noindent
where we introduce the calibration constant $C_1$ to account for the non-squared shape of the emission line profile 
and possible deviations of the filter transmission curves with respect to perfect Heaviside step functions. 

\section{Photometric FWHM}
\label{ap:pw}

Following from Appendix~\ref{ap:pew} we call $FWHM$ the width of the emission line profile and define a narrow band 
\ha~filter (also with a Heaviside step profile) of equivalent width $W_{H\alpha_n} < FWHM$. 
The flux measured by this filter will be

 \begin{equation}
F_{H\alpha_n}= A_{\rm n} + A \left( \frac{W_{\rm n}}{EW}\right)  
 \label{eq:fn}
\end{equation}

\noindent
where $A_{\rm n}=A\times (W_{\rm n}/FWHM)$ is the fraction of line flux measured by the narrow \ha~filter. 
By dividing equation~\ref{eq:fn} by ~\ref{eq:fb} and after some algebra we obtain

\begin{equation}
FWHM=\frac{EW_{ph}}{\left(\frac{EW_{ph}+W_{H\alpha_b}}{W_{H\alpha_n}}\right)\times\left(\frac{F_{H\alpha_n}}{F_{H\alpha_b}} \right)-1} 
\label{eq:w}
\end{equation}
 
\noindent 
As before we define  

 \begin{equation}
FWHM_{ph}=C_{2}~FWHM  
\end{equation}

\noindent
where the constant $C_2$ accounts for the non-squared shape of the emission line profile 
and possible deviations of the filter transmission curves with respect to perfect Heavise step functions.


\bsp	
\label{lastpage}
\end{document}